\documentclass[article,notitlepage, 12pt]{revtex4-1}

\usepackage[utf8]{inputenc}
\usepackage{amssymb,graphicx}
\usepackage{epstopdf}
\usepackage{amsmath,amsfonts}
\usepackage{epsfig}
\begin{document}

\title{Effective phantom dark energy in scalar-tensor gravity}
\author{Anton Saratov}\
\email{saratov@inr.ac.ru}

\affiliation{
Institute for Nuclear Research of
         the Russian Academy of Sciences,\\  60th October Anniversary
  Prospect, 7a, 117312 Moscow, Russia
}

\begin{abstract}
We revisit the problem of phantom behaviour of effective dark energy in scalar-tensor gravity. 
The main focus is  on the 
properties of
the functions defining  the model. We find that
models with the present phantom behavior can be made consistent with all constraints, but one of these functions  
must have rather contrived shape, 
and the initial data must be strongly fine-tuned. Also, the phantom stage must have begun fairly 
recently, at $z\lesssim 1$. All this disfavors the effective phantom behaviour in 
the scalar-tensor gravity.
\end{abstract}

\maketitle

\section{Introduction}
Most models of dark energy 
in the present Universe predict that its effective equation of state satisfies the null 
energy condition (NEC) $w_{eff}=p_{DE}/\rho_{DE}\geq-1$, where $\rho_{DE}$  
and $p_{DE}$ are the effective dark energy density
and pressure, respectively.
However, the observations 
do not rule out that dark energy is
phantom, i.e., it  violates NEC. 
As an example, the 7-year WMAP+BAO+SN data \cite{Komatsu:2010fb} give the following bound on the 
equation of state with time-dependent $w_{eff}$ at $z=0$ :
\begin{equation*}
w_{eff,0}=-0.93\pm0.13 \ (68\% CL) \; ,
\end{equation*}  
which is not entirely inconsistent with $w_{eff, 0} < -1$.
Even though phantom dark energy can be accomodated within General Relativity 
\cite{ArmendarizPicon:2004pm,33,34,galileon}, 
it is legitimate to ask whether effective phantom behavior 
can be obtained in modified gravity theories, such as $f(R)$ or scalar-tensor 
gravity \cite{Bamba:2010bm, Amendola:2006we, od, carrol, 
 SaezGomez:2008uj,  star0}. This question can be
addressed, in particular, within a popular approach~\cite{star0, epol, star2}
employing the reconstruction 
of model parameters from the redshift 
expansion of observables, combined with the experimental constraints on non-GR gravity. 
The main conclusion is that with
current observational data,
the reconstructed DE behaves more or less
like the cosmological constant, but it is still possible to have phantom stage today. 

In this paper we follow a somewhat 
different route and 
ask what sort of the scalar-tensor Lagrangian can lead to effective phantom 
DE at the present epoch
without violating constraints from Solar system and local gravity experiments, 
time-(in)dependence of the gravity constant,
etc.
We also ask whether fine-tuning of initial data is necessary 
and how long the duration of the phantom stage can be in the past.

Our results are somewhat disappoining. We find that
models with the present phantom DE can be made consistent with all constraints, but one of
the functions entering the 
scalar-tensor Lagrangian must have rather specific shape, 
and the initial data must be strongly fine-tuned. Also, the phantom stage must have begun fairly 
recently, at $z\lesssim 1$. 
Before that the scalar field was undistinguishable from quintessence.
All this disfavors the effective phantom behaviour in 
the scalar-tensor gravity.
In fact, as we point out towards the end of this paper,
some of the unpleasant properties we discuss
must be present in scalar-tensor models for effective dark energy irrespectively of whether
it is phantom or not.

This paper is organized as follows. In Section II we present 
the  equations governing the homogeneous cosmological 
evolution in the 
scalar-tensor gravity. 
We recall in Section III the experimental constraints on the non-GR  gravity,
that place bounds on 
the parameters of the theory. 
In Section IV we define the expansion coefficients of the functions 
entering the Lagrangian and reformulate the bounds of Section III in terms of these
coefficients.
In Section V we put together all consistency requirements for the
effective
phantom dark energy today 
and arrive at qualitative understanding of the properties of the functions
defining the theory.
Also, the maximum 
redshift at which the phantom phase could begin is  estimated. 
Section VI contains a numerical example.
We conclude in Section VII.

\section{Homogeneous and isotropic evolution}

By definition, the effective dark energy density and pressure  
$\rho_{eff}$ and $p_{eff}=w_{eff}\rho_{eff}$ are the quantities 
entering the GR-looking evolution equations for the homogeneous and
isotropic Universe,
\begin{align}
3H^{2}&=\rho+\rho_{eff} \label{E:fri1}\\
-2\dot{H}&=\rho+\rho_{eff}(1+w_{eff}) \label{E:fri2} \; 
\end{align}
where $\rho$ and $p=0$ are matter energy density and pressure,
and we set $8\pi G_N = 1$.
Using (\ref{E:fri1}) and (\ref{E:fri2}), one writes
\begin{equation}\label{E:omega}
w_{eff}=-\frac{1}{1-\Omega_{m}}\left(1+\frac{2}{3}\frac{\dot{H}}{H^{2}}\right) ,
\end{equation}
where $\Omega_{m}=\rho/3H^{2}$. We are going to make use of this relation
in the context of the scalar-tensor gravity.
The action of this theory is 
\begin{equation}\label{E:action}
S=\frac{1}{2} \int d^{4}x \sqrt{-g} \left( F(\Phi)R - Z(\Phi) g^{\mu \nu} \partial_{\mu}\Phi \partial_{\nu} \Phi - 2 U(\Phi) \right) + S_{m}(\psi, g_{\mu \nu}),
\end{equation}
(mostly positive signature), where
the action for the usual matter $S_{m}$ 
does not depend on $\Phi$.  One can always redefine the field to have a convenient form of either
$F(\Phi)$ or $Z(\Phi)$. 
We will use the general form of $F(\Phi)$ 
and set 
\[
Z(\Phi)=1.
\]

From the action~(\ref{E:action}) one  obtains the gravitational equations,
\begin{equation}\label{E:ein}
F(\Phi)\left(R_{\mu \nu}-\frac{1}{2}g_{\mu \nu}R \right) =T_{\mu \nu} 
+   \partial_{\mu}\Phi \partial_{\nu} \Phi - \frac{1}{2}g_{\mu \nu}\left(\partial \Phi\right)^{2} 
+\nabla_{\mu}\nabla_{\nu}F(\Phi)  -g_{\mu \nu}\Box F(\Phi)-g_{\mu \nu}U(\Phi)
\end{equation}
and equation of motion for the field $\Phi$,
\begin{equation}\label{E:phi}
\Box\Phi=-\frac{1}{2}\frac{dF}{d\Phi}R+\frac{dU}{d\Phi}.
\end{equation}
Let us specify to the homogeneous, isotropic and spatially flat Universe
with metric
\begin{equation*}
ds^{2}=-dt^{2}+a^{2}(t)dx^i dx^i.
\end{equation*}
Since matter does not interact with $\Phi$, the scale factor $a$ has the
same meaning as in GR.
Using Eqs.~(\ref{E:ein}) and (\ref{E:phi}) one gets the following set of equations:
\begin{align}
3F H^{2}&=\rho+\frac{1}{2}\dot{\Phi}^{2}-3H\dot{F}+U \label{E:fr1j}
\\
-2F\dot{H}&=\rho+\dot{\Phi}^{2}+\ddot{F}-H\dot{F} \label{E:fr2j}\\
\ddot{\Phi}+3H\dot{\Phi}&=3(\dot{H}+2H^{2})\frac{dF}{d\Phi}-\frac{dU}{d\Phi}. \label{E:phj}
\end{align}
The equation for the 
matter density  has the usual form,
\begin{equation*}
\dot{\rho}+3H\rho=0.
\end{equation*}

It is convenient for  futher analysis to switch from the evolution in
time to the evolution in redshift. This can be done by using the relation
\begin{equation*}
\frac{d}{dt}=-H(1+z)\frac{d}{dz}.
\end{equation*}
In this way one obtains from ~(\ref{E:fr1j}), (\ref{E:fr2j}) and (\ref{E:phj}) the evolution
equations in terms of redshift:
 \begin{align}
3F H^{2}&=\rho+H^{2}(1+z)^2\frac{\Phi'^{2}}{2}+3H^{2}(1+z)F'+U \label{E:fr1z}\\
2FHH'&=\rho+H^{2}(1+z)^{2}\Phi'^{2}+H^{2}(1+z)^{2}F''+\nonumber\\&+[(1+z)^{2}HH'+2H^{2}(1+z)]F' \label{E:fr2z}\\
H^{2}(1+z)^{2}\Phi''&+[HH'(1+z)^{2}-2H^{2}(1+z)]\Phi'=3[2H^{2}-HH'(1+z)]\frac{F'}{\Phi'}-\frac{U'}{\Phi'}, \label{E:phz}
\end{align}
where prime denotes $d/dz$. 

\section{Constraints}
The properties of
functions defining the theory are strongly 
constrained by 
local and Solar system experiments. 
This is a major problem for the effective NEC-violating behavior in the
scalar-tensor gravity. 
One important parameter is the Brans--Dicke "constant" $W_{BD}(\Phi)$. It
is straightforward to obtain the expression for  $W_{BD}$ 
in our parametrization by redifining the scalar field.
One finds
\begin{equation}\label{wf}
W_{BD}=\frac{F}{\left(\frac{dF}{d\Phi}\right)^{2}}.
\end{equation}
The lower bound on the
 present value of  $W_{BD}$ is
obtained from the  Kassini experiment \cite{gamma_exp,gamma1,gamma2}. It reads (the subscript $0$ denotes the quantities at the present epoch)
\begin{equation}\label{wbd}
W_{BD, 0}>4\cdot10^{4}.
\end{equation}
A bound of another sort
follows from the experiments on the time dependence of the gravity constant. 
In our case the local gravity constant $G_{loc}$ is given by~\cite{star0}
\begin{equation} \label{E:Ge}
8 \pi G_{loc}=\frac{1}{F}\left(\frac{2F+4(dF/d\Phi)^{2}}{2F+3(dF/d\Phi)^{2}}\right)=\frac{1}{F}\frac{2W_{BD}+4}{2W_{BD}+3}.
\end{equation}  
The experimental constraint on the time evolution of $G_{loc}$ can be found in Refs.~\cite{Williams:2005rv,pit}.  
 For $z=0$ it reads
\begin{equation} \label{Gete}
\left(\frac{\dot{G}_{loc}}{HG_{loc}}\right)_{0}<0.5\cdot10^{-2}.
\end{equation}
We also know that the gravitational constant relevant for cosmology
should not change significantly since Big Bang Nucleosynthesis~\cite{Uzan:2010pm},
\begin{equation} 
\frac{\Delta G_{cosm}}{G_{cosm}}\lesssim 0.1.
\label{mar11-2}
\end{equation}
It is the latter constraint that plays a significant role in our analysis,
see Section V.

\section{Expanding in redshift and $\Phi$}

A convenient way to analyze the evolution at small redshifts is to expand all 
 functions in the Taylor series in redshift $z$. 
On the other hand, we are mainly interested 
in the dependence on $\Phi$, so we will use the mixed expansion.
At $z=0$, without loss of generality we choose
 \begin{equation*}
\Phi_{0}=0 
\end{equation*}
and by definition of the Newton gravity constant we have
\begin{equation*}
F_{0}=1. 
\end{equation*}
Here is our definition of the expansion coefficients:
\begin{align}
F(z)&=1+F_{1}z+\frac{1}{2}F_{2}z^{2}+\frac{1}{6}F_{3}z^{3}...     \label{E:ff}\\
U(z)/3H_{0}^{2}&=\Omega_{U , 0}+U_{1}z             \label{E:uu}\\
H^{2}(z)/H_{0}^{2}&=1+h_{1}z+\frac{1}{2}h_{2}z^{2}+... \label{E:hh1}\\
\Phi'(z)&=\Phi'_{0}z + \frac{1}{2} \Phi_0^{\prime \prime} z^2         \label{E:pp}\\
\rho(z)/3H_{0}^{2}&=\Omega_{m,0}(1+z)^{3}. \label{matrho}
\end{align}
Without loss of generality we take $\Phi'_{0}>0$.
From Eq.~(\ref{E:fr1z}) it is straightforward 
to obtain the relation between the derivatives at the present time,
\begin{equation}\label{phsh}
\Phi_{0}^{\prime 2}=6(1-\Omega_{U,0}-\Omega_{m,0}-F_{1}).
\end{equation}
We will also need $h_{1}$ and $h_{2}$ to obtain the expression for $w_{eff}$. 
From Eqs.~(\ref{E:fr1z}), (\ref{E:fr2z}) we get
 \begin{align}
h_{1}&=\frac{1}{1-\frac{F_{1}}{2}}\left(6-3\Omega_{m,0}-6\Omega_{U,0}-4F_{1}+F_{2}\right), \label{E:h11}\\ 
h_{2}&=\frac{3}{\left(1-\frac{F_{1}}{2}\right)^{2}}[F_{1}\left(\frac{5}{2}F_{1}-3\frac{3}{2}F_{2}-\frac{F_{3}}{12}+4\Omega_{U,0}
+\frac{U_{1}}{2}+\frac{11}{4}\Omega_{m,0}-7 \right)\nonumber\\& +F_{1}^{2}-\frac{3F_{2}}{2}\Omega_{U,0}-\frac{3F_{2}}{4}\Omega_{m,0}+2F_{2}
+\frac{F_{3}}{6}-5\Omega_{U,0}-u_{1}-4\Omega_{m,0}+5].\label{E:h22}
\end{align}
Our main purpose is to understand the behavior of
$F$ and $U$ as functions of the scalar field, so we expand them in $\Phi$: 
 \begin{align}
F=1+f_{1}\Phi+\frac{1}{2}f_{2}\Phi^{2}+\frac{1}{3}f_{3}\Phi^{3}, \label{E:F}\\
U=u_{0}+u_{1}\Phi+\frac{1}{2}u_{2}\Phi^2. \label{E:U}
\end{align}
The relationship between the expansion coefficients entering (\ref{E:ff}) and (\ref{E:F}) is
 \begin{align}
F_{1} &= f_{1}\Phi'_{0} \label{E:F1}\\
F_{2} &= f_{1}\Phi_{0}''+f_{2}\Phi_{0}'^{2} \label{E:F2}\\
F_{3} &=f_{1}\Phi_{0}'''+2f_{2}\Phi_{0}''\Phi_{0}'+f_{3}\Phi'^{3}. \label{E:F3}
\end{align}

We now recall the
constraint on $W_{BD,0}$, Eq.~(\ref{wbd}), and make use of Eq.~(\ref{wf}).
With our normalization $F_{0}=1$, we get very strong upper bound on $f_{1}$:
\begin{equation}
|f_{1}| <0.5\cdot10^{-2}.
\label{mar11-1}
\end{equation}
This means that the field $\Phi$ is presently near the extremum of the 
funcion $F(\Phi)$. Such a conclusion appears inevitable in modified gravity, see, e.g., Ref.~\cite{carrol}.

It follows from Eq. ~(\ref{phsh}) that
$\Phi_{0}^{\prime 2}\lesssim 1$, so Eq.~(\ref{E:F1}) implies that $F_{1}$
is also small,
\begin{equation} \label{f1r}
|F_{1}|\lesssim 10^{-2}.
\end{equation}
This suggests that we can neglect terms with the first derivative of $F$ in the
analysis of the present epoch. We note in passing that 
a small value of $F'(z=0)$ could have been anticipated, since the GR tests are very precise,
 and only a slight 
deviation from GR can be tolerated today.
 
Using ~(\ref{phsh}) and neglecting the term with
$F_{1}$, we obtain for the present value of the field derivative with respect to redshift
\begin{equation}\label{ph2}
\Phi_{0}^{\prime 2}=6\left(1-\Omega_{U,0}-\Omega_{m,0}\right).
\end{equation}
This simple relation will be instrumental in what follows.

\section{Effective phantom behavior}
Now we use the expression (\ref{E:omega}) 
for $w_{eff,0}$ to find out which parametrs can be responsible for the
 effective phantom behavior. 
Making use of Eqs.~(\ref{E:F1}),(\ref{E:F2}) and (\ref{f1r}) we get
\begin{equation} \label{ww0}
1+w_{eff,0}=\frac{f_{2}(\Phi'_{0})^{2}+6(1-\Omega_{m,0}-\Omega_{U,0})+f_{1}\Phi''_{0}}{3(1-\Omega_{m,0})}.
\end{equation}
By extracting the second derivative of the field from Eq.~(\ref{E:phz}), 
we find
\begin{equation*}
\Phi_{0}''=\left( 2-\frac{h_{1}}{2} \right)\Phi_{0}'-U_{1}.
\end{equation*}
So, there are essentially
two parameters that could yield $1 + w_{eff, 0} < 0$, namely, $f_2$ and $U_1$.
We begin our discussion with the latter.

The possible contribution of the potential to the phantom effective equation of state comes 
from the third term in (\ref{ww0}) (the second term in the 
numerator is positive in virtue of Eq.~(\ref{ph2}))
and is given by
\begin{equation}
\Delta_{U}(1+w_{eff,0})=-f_{1}\frac{U_{1}}{3(1-\Omega_{m,0})}.
\label{mar10-1}
\end{equation}
It is strongly supressed by small $f_{1}$, so this contribution can be sizeable only if
the potential $U(\Phi)$ is very steep today.
However, steep potential would lead to the rapid  acceleration 
of the scalar field, so the phantom phase 
would  be very short in the past. 
Furthermore, the fast evolution of the scalar field together with large
$u_1 = dU/d\Phi (z=0)$ would imply rapid change in time of the effective
dark energy density. To elaborate on the latter point, let us consider the parameter
\[
w_1 = \frac{d w_{eff}}{dz} (z=0).
\]
Observationally, $|w_1|$ is not large: the WMAP analysis \cite{Komatsu:2010fb} gives
$-0.31 < w_1 < 1.12$.
On the other hand, making use of Eq.~(\ref{E:omega}) and neglecting the terms with $F_1$
we obtain
\begin{equation}\label{ww1} 
w_{1}=w_{1,A}+w_{1,B},
\end{equation}
where $w_{1,A}$ and $w_{1,B}$ are:
\begin{align}
w_{1,A}&=\frac{1}{3(1-\Omega_{m,0})}(F_3 - 6u_1), \label{ww11}\\
w_{1,B}&=\frac{1}{3(1-\Omega_{m,0})^{2}}\left[(1+5\Omega_{m,0})\left(F_{2}-6\Omega_{U,0}-3\Omega_{m,0}+6\right)-9\Omega_{m,0}\right]+\nonumber\\
&+\frac{1}{3(1-\Omega_{m,0})}\left[12F_{2}+\frac{3F_{2}^{2}}{2}+9F_{2}\Omega_{U,0}-\frac{9}{2}F_{2}\Omega_{m,0}
-30\Omega_{U,0}\right.\nonumber\\
& \left.+24(1-\Omega_{m,0})+6\phantom{\frac{1}{1}}\right]-\frac{1}{3(1-\Omega_{m,0})^{2}}\left(-F_{2}+6\Omega_{U,0}+3\Omega_{m,0}-6\right)^{2}. \label{ww12}
\end{align}
The parameter $F_2$ cannot be very large, see below, so for large $u_1 = U_1/\Phi_0^\prime$
the value of $w_1$ is controlled by $w_{1,A}$ term in the expression~(\ref{ww1}).
To have sizeable contribution (\ref{mar10-1}) and at the same time
satisfy the observational constraint on $w_1$, one would need
the cancellation between $F_3$ and $6u_1$, which in turn would require strong fine-tuning.
Barring this possibility, we arrive at the conclusion that $|f_1 U_1| \ll 1$,
so the contribution (\ref{mar10-1}) is very small.
From now on we neglect it.

The remaining terms in Eq.~(\ref{ww0}) 
can be simplified by using ~(\ref{ph2}):
\begin{equation} \label{ww}
1+w_{eff,0}=\frac{(1+f_{2})(\Phi'_{0})^{2}}{3(1-\Omega_{m,0})}.
\end{equation}
Thus, the phantom behaviour today is controlled entirely by $f_{2}=d^{2} F/d \Phi^{2}(z=0)$. 
To have $w_{eff}<-1$ at the present time, one requires that
\begin{equation*}
f_2 < -1.
\end{equation*}
Together with the bound (\ref{mar11-1}), this implies that 
today the field $\Phi$ must be close to a relatively sharp
maximum of the function $F(\Phi)$. Clearly,
such a special state requires fine-tuning of both the function $F(\Phi)$ and initial conditions in the theory.

\begin{figure}[h]
\center{\includegraphics[width=90mm]{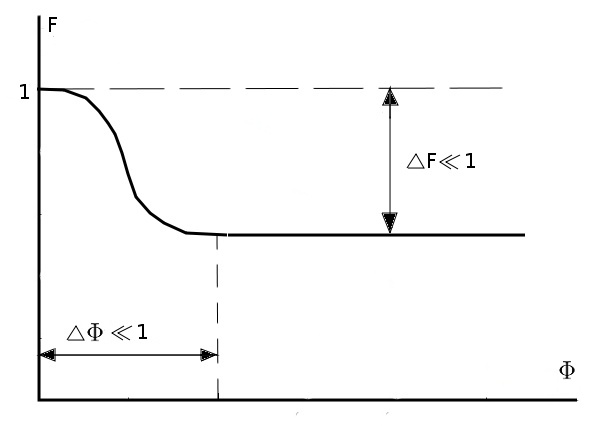}\\}
\caption{The shape of $F(\Phi)$. The present value of $\Phi$ must be near the maximum.}
\label{ris:image}
\end{figure}

We continue the discussion of the shape of $F(\Phi)$ and recall the constraints on the time-dependence of
the gravity constant. Given the small value of $F_1$ and large value of $W_{BD, 0}$, the constraint
(\ref{Gete}) is not hard to satisfy; note that this is in contrast to Ref.~\cite{babichev}. 
Much less trivial is the fact that the ``cosmological''
gravity constant  
has not changed much since BBN. It is clear from (\ref{E:fr1j}) that $G_{cosm}(\Phi)$ is simply equal to $F^{-1}(\Phi)$. Barring strong cancellations, the constraint (\ref{mar11-2}) together with
Eq.~(\ref{E:fr1z}) imply that the variation of $F(\Phi)$ has been small since BBN, 
\begin{equation}
\Delta F \lesssim 0.1
\label{mar11-5}
\end{equation}
and that $F^\prime$ is small at large $z$. Thus, the function $F(\Phi)$ must have the shape shown in
Fig.~\ref{ris:image}.

Let us now estimate the range of redshifts in which the dynamics of $F$ is non-trivial, and the phantom
effective equation of state can be realized. We do this by requiring that the value of $F$ does not change much
during this period. Since $F_1$ is small, the second term in the expansion of $F$ in redshift
is relevant, and the estimate for maximum redshift is found from
\[
\frac{1}{2} F_2 z_{max}^2 \lesssim \Delta F,
\]
where the bound on $\Delta F$ is given in Eq.~(\ref{mar11-5}). Making use of Eq.~(\ref{E:F2})
we get
\begin{equation}
\frac{1}{2}(|f_{2}|\Phi'_{0})^{2}z_{max}^{2}\lesssim\Delta F .
\label{g2} 
\end{equation}
Let us denote by $\epsilon$ the deviation of $w_{eff}$ from $-1$ today:
\begin{equation*}
1+w_{eff,0}=-\epsilon.
\end{equation*}
Using (\ref{ww})  we get the estimate
$(1+f_{2})(\Phi'_{0})^{2}\approx 3 \epsilon$, and hence from~(\ref{g2}) we find
\begin{equation} \label{zmax1}
z_{max}^{2}\lesssim \frac{2\Delta F}{3\epsilon}\frac{|1+f_{2}|}{|f_{2}|}.
\end{equation}
For reasonably strong phantom behavior (i.e., not very small $\epsilon$), $z_{\max}$ is fairly
small; roughly speaking, $z_{max} \lesssim 1$. 
Note that similar result has been obtained within the reconstruction approach~\cite{carrol,star2, star1}.
At larger redshfts, $F$ is frozen out, and the scalar field
$\Phi$ reduces to quintessence.

\section{Numerical example}

\begin{figure}[h!]
\center{\includegraphics[width=90mm]{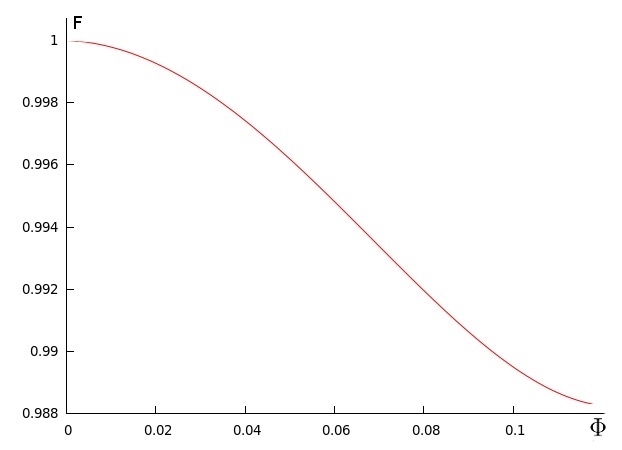}\\}
\caption{$F(\Phi)$ in the numerical example.}
\label{image3}
\end{figure}

Let us give a concrete example of a model with effective phantom behavior
today and in the recent past. We note that the relatively large value
of the parameter $\epsilon = - (1+w_{eff, 0})$ is obtained for fairly large
$\Phi_0^\prime$, otherwise the maximum of $F$ must be very sharp (i.e., $f_2$
must be large), see Eq.~(\ref{ww}). So, we take, somewhat arbitrarily, $\Phi_0^\prime = 0.3$.
We choose $\Omega_{m,0}=0.25$, in rough agreement with observations, then Eq.~(\ref{E:fr1z})
with $F_1 \ll 1$ gives $\Omega_{U,0}=0.73$. Fairly strong phantom behavior, $\epsilon =0.1$,
is obtained with $f_2 = -1.75$. The estimate (\ref{zmax1}) then gives $z_{max} \sim 0.2$.
To satisfy all these requirements, we choose $F(\Phi)$ as shown in Fig.~\ref{image3}. 

\begin{figure}[h!]
\center{\includegraphics[width=90mm]{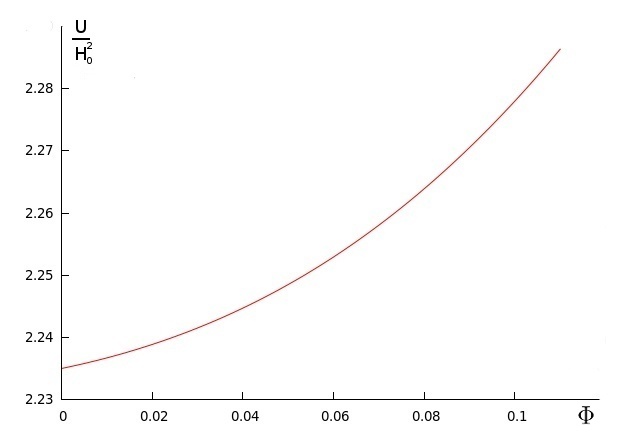}\\}
\caption{$U(\Phi)$ in the numerical example}
\label{image6}
\end{figure}

The shape of the potential $U(\Phi)$ is not constrained particularly strongly; our choice is shown in
Fig.~\ref{image6}. 

\begin{figure}[h!]
\center{\includegraphics[width=90mm]{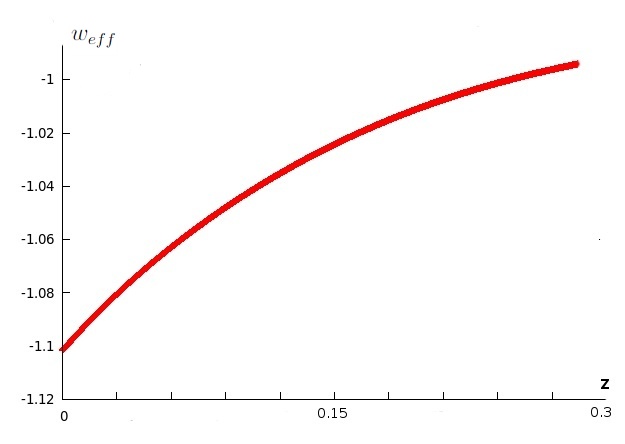}\\}
\caption{$w_{eff}$ as function of redshift.}
\label{image2}
\end{figure}

With this choice, the effective equation of state depends on redshift as shown in Fig.~\ref{image2}.
As expected, $w_{eff}$ rapidly tends to $-1$ as $z$ increases, and the field $\Phi$ becomes
indistiguishable from quintessence at $z \gtrsim 0.2$. The field value does not change much:
the change from redshift 0.2 to the present epoch is about  $\Delta \Phi \approx0.05$. 

We have constructed a number of other examples satisfying the constraints of
Section III; all of these examples have similar properties.

\begin{figure}[h!]
\center{\includegraphics[width=90mm]{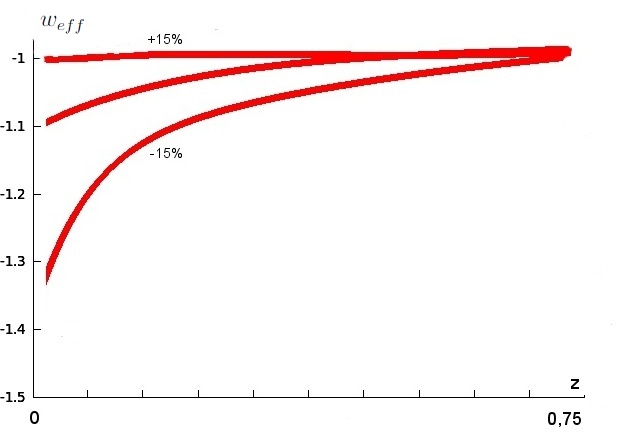}\\}
\caption{$w_{eff}$ for initial conditions for $\Phi$ deviating by $\pm 15\%$ from the central value.}
\label{image8}
\end{figure}

As we already pointed out,
effective phantom 
dark energy requires both the special form of the function $F(\Phi)$ and  fine-tuning of the initial value of the 
field $\Phi$. To see the latter property explicitly, let us take the same functions $F(\Phi)$ and $U(\Phi)$
as before and
consider the evolution from redshift $z=0.75$ to $z=0$ for different initial values of the field. 
If we vary the initial condition for $\Phi$ within 15\% interval around the central value
yielding Fig.~\ref{image2} (without varying the initial velocity $\dot{\Phi}$, for the sake of arument), 
the evolution changes considerably. In particular, the effective equation of state
is as shown in Fig.~\ref{image8}.

Such a behaviour is not unexpected. The right choice of the
initial value of the field ensures that the phantom phase begins in just right time, at some rather small redshift.
For different initial values, the onset of the  phantom behavior 
occurs at ``wrong'' redshifts, so one either has too large deviation of $w_{eff,0}$ from $-1$, or no deviation at
all.

\section{Conclusions}
In this paper we revisited the question of the
possibility of the present phantom phase in scalar-tensor gravity.
We have seen that it 
is possible to obtain and control 
effective phantom behavior even in simple  scalar-tensor models, but
this requires a lot of fine-tuning. First, the large present value of the
Brans--Dicke parameter is obtained only if the scalar field $\Phi$ is presently
near the extremum of the function $F$ determining the gravity constant.
This is consistent with observable phantom property only if this extermum is a
sharp maximum. Second, the small variation of the gravity constant since BBN
requires that $F$ flattens out at fairly low redshift. Finally, the whole picture is
consistent with observations only for fine-tuned initial data.

We conclude by noting that some of the unpleasant properties discussed in this paper
must be present in scalar-tensor models for effective dark energy irrespectively of whether
it is phantom or not. This remark applies, in particular, to the contrived shape of the
function $F$ and fine-tuning of initial conditions. Indeed, the fact that $dF/d \Phi$
must be small today does not rely on the assumption of the phantom behavior. 
Furthermore, most of the
analysis in Section V goes through provided that $|d^2 F /d\Phi^2|$ is large enough at the
present epoch, while the case $|d^2 F/d\Phi^2| \ll 1$ corresponds to
quintessence rather than genuine scalar-tensor gravity. All this makes scalar-tensor theory
rather unlikely candidate for explaining the accelerated expansion of the Universe.


\end{document}